\documentclass[%
 reprint,
 amsmath,amssymb,
 aps,
]{revtex4-2}

\usepackage{graphicx}
\usepackage{dcolumn}
\usepackage{bm}
\usepackage{siunitx}
\DeclareSIUnit{\sqrthertz}{\sqrt{\unit{\hertz}}}

\begin{document}

\preprint{APS/123-QED}

\title{A 3D passive ring gyroscope for seismology}

\author{Thomas Gereons}
\email{tgereons@uni-bonn.de}
\author{Jannik Zenner}
\author{Thorsten Groh}
\author{Simon Stellmer}
\affiliation{Physikalisches Institut, Universität Bonn, Nussallee 12, 53115 Bonn, Germany}

\date{\today}

\begin{abstract}
In seismology and related fields, the measurement of rotation in all three spatial dimensions is essential to complement the observation of translations. Access to all six degrees of freedom allows for full reconstruction of seismic wavefields and improves the understanding of complex ground motion during seismic events. In this regard, Sagnac interferometers in the form of large \emph{active} ring laser systems have demonstrated remarkable performance. So-called \emph{passive} ring gyroscopes offer the potential to bypass some of the limitations of active ring lasers and could represent a promising complement to existing sensor technology. Here, we present a prototype of a transportable three-dimensional free-space passive ring gyroscope, reaching a sensitivity in the \SI{}{\micro \radian / \second / \sqrthertz} regime in all spatial dimensions. We demonstrate the sensor performance by reconstructing the rotational components of a simulated seismic event.
\end{abstract}

\maketitle

\section{\label{sec:level1}Introduction}

Highly sensitive rotational sensors play a fundamental role in a wide range of scientific and technological applications, including navigation \cite{Ghufran2024}, seismology \cite{Igel2005,Igel2011,Igel2021} and geodesy \cite{U.Schreiber2013,A.Gebauer2020,Schreiber2023,Schreiber2025}. In the geosciences, rotation measurements provide direct access to rotational ground motion generated by seismic wavefields, as well as to variations in the Earth's rotation. These observations complement translational measurements and enable a more detailed physical description of geodynamic processes. Ring laser gyroscopes (RLGs) \cite{Schreiber2023Book}, which operate on the basis of the Sagnac effect \cite{G.Sagnac1914}, are among the most precise instruments for measuring rotation rates. When two coherent light beams propagate in opposite directions along a closed optical path in a rotating reference frame, they accumulate a phase difference proportional to the enclosed area $A$ and the rotation rate $\mathbf{\Omega}$. In an optical ring cavity of perimeter $P$, this phase shift appears as a beat frequency $\delta f$ between the counter-propagating modes, called the Sagnac frequency,
\begin{equation}
\delta f = \frac{4A}{\lambda P}\,\mathbf{\Omega}\cdot \mathbf{n},
\label{eqn:sag}
\end{equation}
where $\lambda$ is the wavelength and the scalar product projects the rotation rate onto the cavity normal $\mathbf{n}$ . The scale factor $4A/(\lambda P)$ is exclusively dependent on the design of the optical cavity. Optical gyroscopes based on the Sagnac effect can be broadly divided into active and passive configurations. Active RLGs \cite{Schreiber2003,Virgilio2019,brotzer2025,Zenner25} use a gain medium, typically He-Ne, inside the optical cavity to produce counter-propagating laser modes. They offer excellent long-term stability and narrow linewidths, but suffer from gain degradation, multi- and split-mode operation, and backscatter-induced mode coupling \cite{U.Schreiber2013,brotzer2025,Chen2025}.
Passive ring gyroscopes (PRGs), by contrast, decouple the laser source from the sensing cavity and probe the resonance using an external laser. These systems are commonly realized as free-space resonant cavities \cite{Ezekiel1977,Korth_2016,Liu:19,Feng2023,Chen2025,Zenner:26}. Fiber-optic gyroscopes (FOGs) \cite{Yuan2020,Izgi2021,Chung2022,Huang:24}, while also based on the Sagnac effect, constitute a distinct interferometric architecture employing coiled optical fibers rather than resonant cavities. Fiber-optic gyroscopes are compact but limited by Rayleigh backscattering and thermally induced refractive-index fluctuations (Shupe effect) \cite{U.Schreiber2013}. Free-space passive cavities suppress many of these noise sources and allow independent control of optical power \cite{Chen2025}. Since, in the passive approach, the probe power can be freely adjusted, the extracted optical power can exceed the power extracted from an active RLG by many orders of magnitude. As this directly translates into the signal-to-noise ratio at the detecting photodiode, passive systems can provide high sensitivity, particularly for seismological applications, where sensitivity on timescales down to milliseconds is more critical than long-term stability \cite{Schreiber2023Book,Brotzer2023}.

Over the past decades, large-scale active RLGs have achieved remarkable performance. Instruments such as the G-ring at the Geodetic Observatory Wettzell \cite{Schreiber2023,Schreiber2025} or the GINGERINO system located at the Laboratori Nazionali del Gran Sasso (LNGS) \cite{Virgilio2019,Virgilio2024} have reached significant sensitivities, allowing to determine Earth's rotation rate at a fractional uncertainty of $10^{-8}$. These measurements enabled the resolution of geophysical phenomena such as polar motion \cite{U.Schreiber2013,Schreiber2023Book}, length-of-day variations \cite{Schreiber2023}, and of the precession and nutation of the Earth's rotation axis \cite{Schreiber2025}. 

In seismology, earthquake-induced motions excite not only translational but also rotational components of ground motion \cite{Simonelli2018}. A complete reconstruction of the seismic wavefield therefore requires a six degrees of freedom (6 DoF) measurement \cite{brotzerphd}. Consequently, traditional translational sensing techniques must be complemented by precise rotational measurements, giving rise to the field of rotational seismology. The required rotational sensor sensitivity is on the order of \si{\nano\radian\per\second}, while covering a frequency bandwidth from \SI{10}{\milli\hertz} to \SI{1}{\kilo\hertz} \cite{Schreiber2023Book}.

In recent years, fiber-optic gyroscopes have enabled 6 DoF measurements of seismic events in tectonically active regions such as volcanoes \cite{Capezzuto2024,BrotzerIgel2025}, providing new insights into previously unobserved ground motion characteristics. However, the sensitivity achieved by large-scale active ring laser gyroscopes remains unmatched. Combining this high sensitivity with full rotational vector measurements offers the possibility to approach the rotational background noise level predicted by the Rotational Low Noise Model (RLMN) \cite{Brotzer2023}. 

The large three-dimensional array of active RLGs ROMY \cite{Igel2021,brotzer2025,brotzerphd} near Munich, Germany, has demonstrated the capability of this approach, enabling the reconstruction of full seismic wavefields from rotational data at sensitivities close to the RLMN. This success highlights the common interest in combining the transportability of fiber-optic gyroscopes with the unmatched sensitivity of ring laser gyroscopes. It further emphasizes the need for compact yet highly sensitive rotational instruments capable of operating in challenging environments.

Our work focuses on compact and transportable three-dimensional free-space PRG systems capable of achieving sensitivities comparable to established large-scale installations. Such instruments would enable local full-vector reconstruction of seismic wavefields, improve seismic source characterization, and facilitate rotational measurements at geographically diverse sites. Transportable systems also constitute a step toward distributed international arrays for rotational ground-motion monitoring. Unlike permanently installed stations, portable instruments enable flexible deployment in regions of interest, including areas with sparse instrumentation or temporary measurement campaigns following significant seismic events. The collection of such diverse data has the potential to significantly improve the spatial resolution of rotational measurements and enhance our understanding of seismic wavefields. In this paper, we present a prototype of a transportable three-dimensional free-space  PRG and demonstrate its performance in seismological measurements.

\section{Setup}
\begin{figure*}[t]
    \centering
    \includegraphics[width=\linewidth]{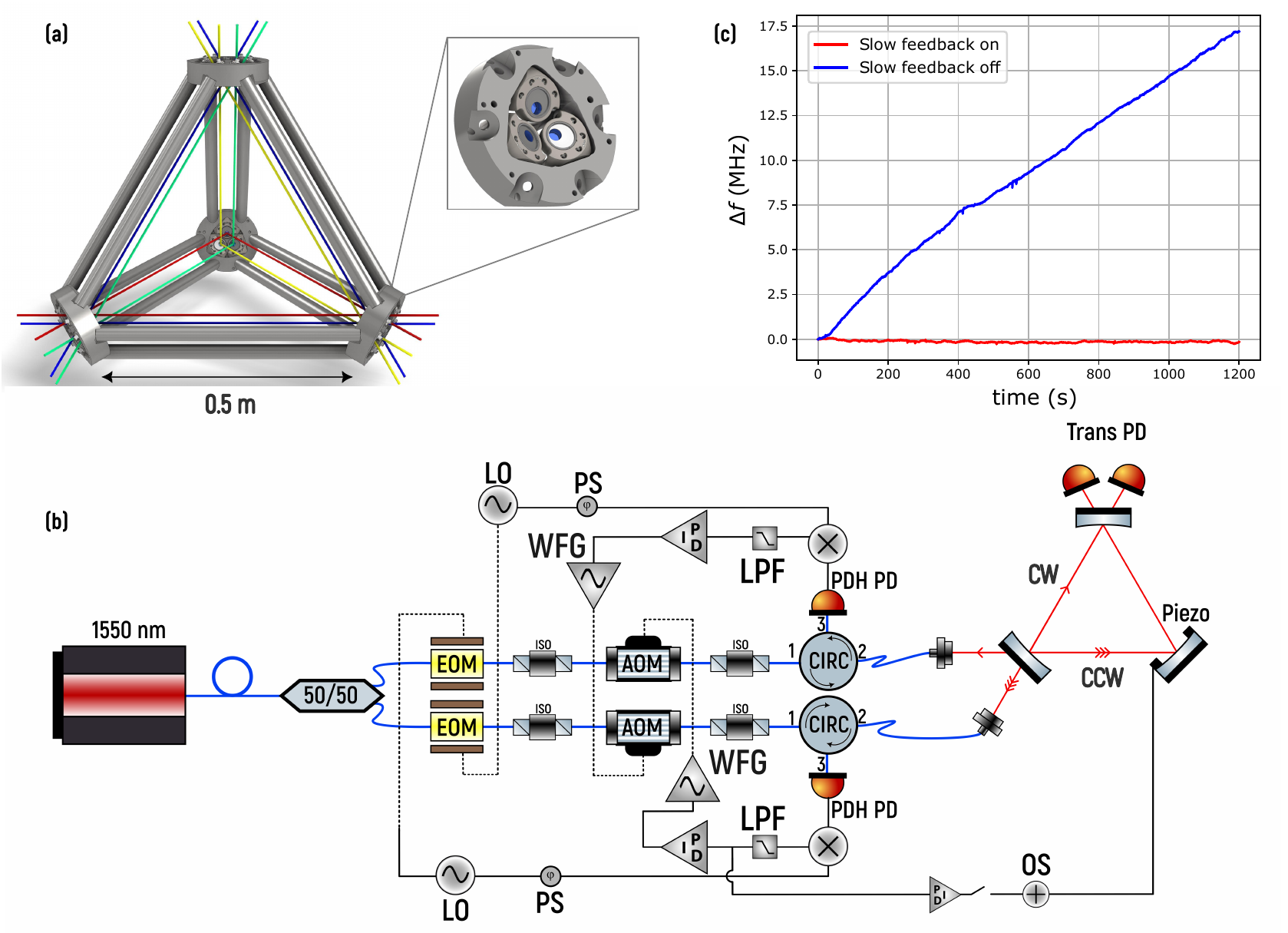}
    \caption{(a) Mechanical design of the three-dimensional passive ring gyroscope as a tetrahedral geometry. The inset shows a close-up view of one of the corner boxes holding the cavity mirrors. Beam paths of the four cavities are shown in different colors. (b) Optical setup for one of the four cavities. Red lines represent laser beams propagating in free-space, blue connections represent polarization maintaining fiber connections, and black lines represent electronic connections. Abbreviations: WFG for waveform generator, PD for photodiode, ISO for isolator, CIRC for circulator, LPF for low-pass filter, LO for local oscillator, OS for adjustable offset, PS for phase delay, PID for proportional-integral-derivative controller, EOM for electro-optic modulator, and AOM for acoustic-optic modulator. The circulator directs light exclusively from port 1 to port 2 and from port 2 to port 3. All other connections are strongly suppressed by 50 dB. Identical configurations are used for the other cavities, apart from the horizontal one, where the laser is directly stabilized to the counter-clockwise direction and only the clockwise direction is locked using the radio-frequency drive of the AOM. (c) Long-term frequency drift $\Delta f$ while stabilizing the radio-frequency drives of one cavity with slow-feedback engaged (red) and without slow-feedback engaged (blue).}
    \label{fig:Setup}
\end{figure*}
The experiment is based on a three-dimensional PRG realized in a tetrahedral geometry. Four corner boxes are connected by twelve stainless-steel rods, forming four triangular ring cavities with high mechanical rigidity, as pictured in Fig.~\ref{fig:Setup}(a). The schematic shown in Fig.~\ref{fig:Setup}(a) is mounted on three stainless steel 1-inch posts, positioned \SI{100}{\milli\meter} above the optical table. Each cavity has a side length of $L = 0.5\,\mathrm{m}$, resulting in a perimeter of $P = $ \SI{1.5}{\meter} and a corresponding free spectral range of $\Delta\nu_{\mathrm{FSR}} = c/P =$ \SI{200}{\mega \hertz}. The cavities are formed by plano-concave mirrors with a radius of curvature of $R_{\mathrm{roc}} = -0.5\,\mathrm{m}$ and a transmission of $T \approx 30\,\mathrm{ppm}$. The finesse $\mathcal{F}$ of the four cavities ranges between 25\,000 and 45\,000. The mirrors are mounted in precision kinematic holders integrated into the corner blocks. The tetrahedral layout provides four independent ring cavities, enabling three-dimensional rotation measurements as well as internal consistency checks via comparison of individual cavity signals. By alternating the three cavities used for the three-dimensional reconstruction, systematic differences between the individual cavities can be investigated, providing a valuable tool for analyzing, understanding, and optimizing the overall system.\\

The optical layout of the experiment is depicted in Fig.~\ref{fig:Setup}(b). A commercial single-frequency fiber laser (NKT Koheras ADJUSTIK E15 \cite{E15Laser}), operating at $1550\,\mathrm{nm}$ with an output power of $40\,\mathrm{mW}$ and a linewidth below $100\,\mathrm{Hz}$, is used as a source. The laser light is split into clockwise (CW) and counter-clockwise (CCW) propagation directions for each cavity using a 50:50 fiber splitter. Phase modulation sidebands required for Pound-Drever-Hall (PDH) stabilization \cite{Black2001} are generated by fiber-coupled electro-optic modulators (EOM) placed in each optical path. Here, the CW direction is phase-modulated at 4.0 MHz while the CCW direction is modulated at 1.6 MHz. After phase modulation, the light is frequency-shifted by fiber-coupled acousto-optic modulators (AOM) by about 80 MHz, allowing independent control of the optical frequencies in the two directions. Optical isolators suppress back-reflections, while fiber circulators separate incident and reflected fields for detection of the cavity reflection signals. The light is coupled into free space using adjustable fiber collimators mounted in five-axis mirror mounts and mode-matched to the cavities with a beam waist of approximately $w_0=$\SI{355}{\micro\meter} located at the midpoint between two cavity mirrors.\\

Frequency stabilization of the optical cavities is implemented using the PDH locking technique, with the control loop executed on a field-programmable gate array (FPGA). The reflected light from the cavity input mirror is detected with a fast photodiode and mixed with a local oscillator (LO) at the phase-modulation frequency. A tunable phase delay maximizes the slope of the demodulated error signal, which is low-pass filtered to obtain a signal proportional to the detuning between laser and cavity resonance. This signal is processed by a proportional-integral-derivative controller (PID) that feeds back to the radio-frequency drive of the AOMs, denoted as WFG in Fig.~\ref{fig:Setup}(b). The RF signal fed to the AOM consists of the nominal operating frequency (80 MHz in this case), which is additionally frequency-modulated proportionally to the PID output. This approach allows precise and parallel locking of each propagation direction to its respective cavity resonance in a compact fiber based setup. This further improves the transportability as no bulky free space optics setup is needed.\\

Since only one circulation can act as the primary frequency reference for the laser, which is the CCW direction of the horizontal cavity in this case, the remaining angled cavities must be tuned into resonance by adjusting their optical path lengths. Small variations in cavity perimeter arising from machining tolerances, alignment offsets or environmental influences well exceed the 5 MHz frequency tuning range of the AOMs. To enable simultaneous operation of all cavities, one mirror in each angled cavity is mounted on a piezoelectric actuator, providing a displacement range of approximately \SI{3}{\micro\meter}, corresponding to a resonance tuning range of about \SI{400}{M\hertz}, i.e., approximately two free spectral ranges. A slow feedback loop acting on these actuators strongly suppresses low-frequency perimeter drifts and enables stable simultaneous operation of all cavities, reducing the average drift from \SI{1}{\mega\hertz/\minute} to \SI{8}{\kilo\hertz/\minute}, as shown in Fig.~\ref{fig:Setup}(c).\\

Rotation of the cavity system induces a Sagnac frequency shift $\delta f$ between the CW and CCW modes, according to Eq.~\ref{eqn:sag}. In the present implementation, the rotation signal is obtained electronically from the difference between the radio-frequency drive signals applied to the AOMs in the two propagation directions. The frequency of the laser light in the CW and CCW modes is determined by the respective AOM, which corresponds to the applied RF signal, that is adjusted proportional to the PID output voltage. Consequently, the frequency difference between the two RF signals is proportional to the applied rotation and therefore provides a direct measure of the Sagnac frequency. To measure this difference, directional couplers are used to mix a fraction of the RF signals driving the AOMs. The resulting signal is then low-pass filtered with a cutoff frequency of 10 kHz in order to suppress high-frequency components, yielding a beat signal at the Sagnac frequency without relying on the optical beat detection. 

\section{Results}
To reconstruct the three-dimensional rotation vector, a coordinate system is defined with respect to the optical table: the $x$-axis along the long side of the table, the $y$-axis along the short side, and the $z$-axis pointing upwards. Each cavity measures only the projection of the rotation vector onto its surface normal, since the Sagnac effect is sensitive exclusively to rotations perpendicular to the cavity plane. The sensitivity of each cavity to rotations about the table axes is therefore determined by its normal vector $\mathbf{n_i}$. These normals are obtained from the cross product of two spanning vectors of the respective cavity plane and are subsequently normalized. Because the rotation vector has three independent components, measurements from three non-coplanar cavities $i=0,1,2$ are sufficient to reconstruct the full three-dimensional rotation. The fourth cavity can additionally be used to extract the rotation rates from an overconstrained set of equations, thereby reducing statistical uncertainties and further lowering the intrinsic noise level of the system. However, this approach has not been implemented to date due to performance limitations in one of the cavities. Since each cavity measures the projection of the rotation vector $\mathbf{\Omega} = \mathbf{\Omega_{xyz}}$ onto its normal, the measured cavity rotation rates $\mathbf{\Omega}_{0,1,2}$ are related by
\begin{equation}
\mathbf{N}\cdot\mathbf{\Omega}_{xyz}=\mathbf{\Omega}_{0,1,2}\:;\:\:\mathbf{N} = \begin{pmatrix}
\dots \mathbf{n_0} \dots \\
\dots \mathbf{n_1} \dots \\
\dots \mathbf{n_2} \dots 
\end{pmatrix},
\end{equation}
where $\mathbf{N}$ is the transformation matrix formed from the normal vectors $\mathbf{n_0,n_1,n_2}$ of the selected cavities. Inverting this relation yields the rotation components in the table frame
\begin{equation}
\mathbf{\Omega}_{xyz}=\mathbf{N}^{-1}\cdot\mathbf{\Omega}_{0,1,2}.
\label{eqn:transformation}
\end{equation}

\begin{figure*}[t]
    \centering
    \includegraphics[width=\linewidth]{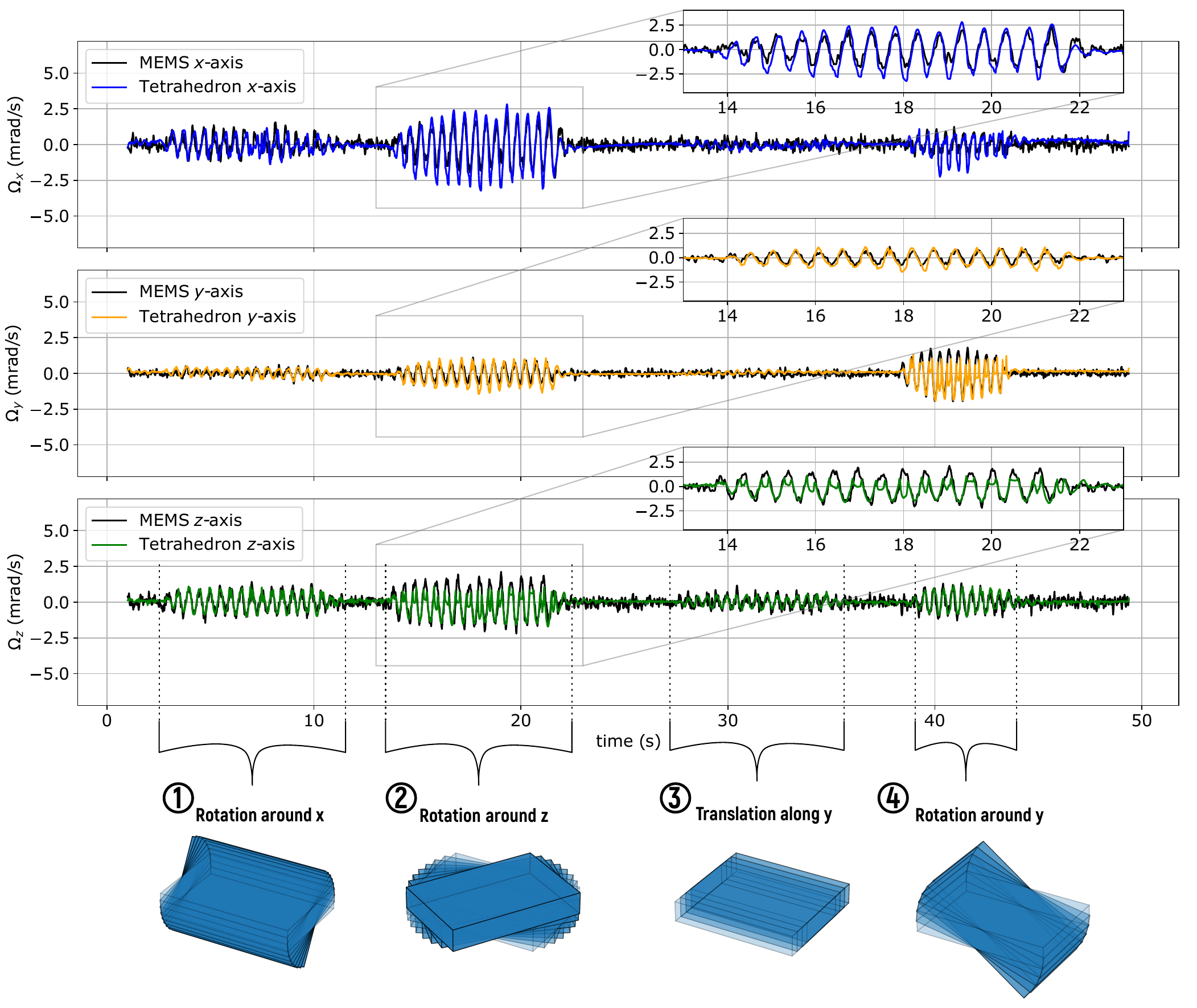}
    \caption{Reconstructed three dimensional rotation rates, denoted as $\Omega_x$ (blue), $\Omega_y$ (orange), and $\Omega_z$ (green), while simulating a seismic activity by shaking the optical table following the illustrated sequence. The excitations are approximately applied over the following time intervals: \SIrange{2.5}{11}{\second} (rotation around $x$), \SIrange{13}{22}{\second} (rotation around $z$), \SIrange{27}{36}{\second} (translation along $y$), and \SIrange{39}{44}{\second} (rotation around $y$). The recording of a commercial MEMS-based gyroscope (\textit{ADIS16465-1} Analog Devices) is given for comparison (black).}
    \label{fig:3Dmeas}
\end{figure*}
For validation, a MEMS gyroscope (\textit{ADIS16465-1} Analog Devices \cite{ADIS}) is mounted on the optical table with its axes aligned to the defined table coordinate system. During the measurement, the laser is PDH-locked to the CCW direction of the horizontal cavity, while the remaining directions are stabilized via the corresponding AOM frequencies. Once all eight branches are locked, the Sagnac beat notes $\delta f$ of all cavities are recorded simultaneously together with the MEMS gyroscope output. Controlled tilting of the optical table about each axis is applied to generate well-defined rotation signals and simulate seismic events, as illustrated by the sequence in Fig.~\ref{fig:3Dmeas}. The measured Sagnac frequency shifts are converted into rotation rates using the calibrated scale factor, resulting in the colored traces of Fig.~\ref{fig:3Dmeas}. The reconstructed rotation vector shows good agreement with the MEMS reference, as exemplified in the inset figures. Note that wavering motion of the optical table induces substantial cross-coupling between the principal axes. Similarly, a pure translation (step 3 in Fig.~\ref{fig:3Dmeas}) also induces mild residual rotations, as measured by the sensors. Hence, we believe that any observed coupling between the rotational signals of the individual axes arises from the manner in which the table is driven to simulate a seismic event.

Nevertheless, the three-dimensional RLG accurately reconstructs the instantaneous 3D rotation rate, demonstrating its performance as a prototype system.\\
\begin{figure*}[t]
    \centering
    \includegraphics[width=\linewidth]{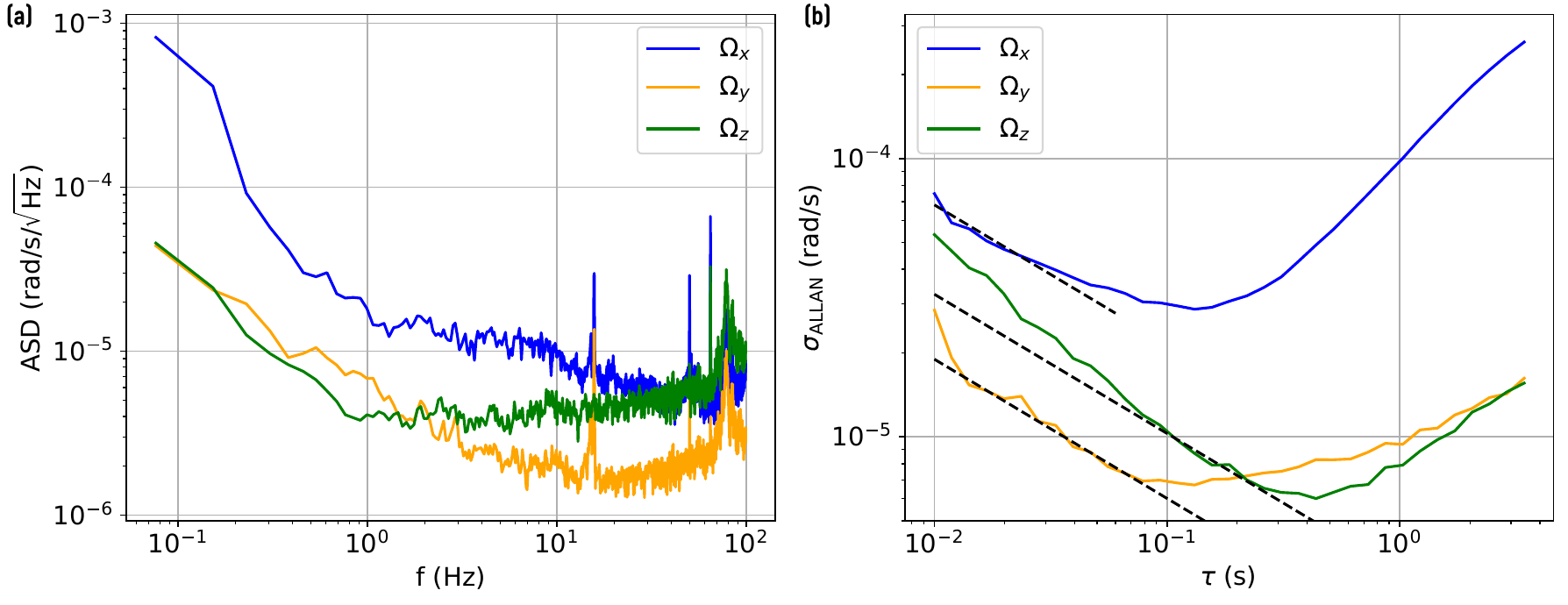}
    \caption{(a) Amplitude spectral densities of the reconstructed rotation rate from 0.01 Hz to 100 Hz for all three spatial dimensions, denoted as $\Omega_x$ (blue), $\Omega_y$ (orange), and $\Omega_z$ (green). (b) Allan deviation of the reconstructed rotation rate for all spatial dimensions using the same color code, shown for integration times from 0.01\,s to 4\,s. The white-noise-limited regime is indicated by black dashed lines by fitting a function of the form $\sigma_{\mathrm{wn}} = \frac{A}{\sqrt{\tau}}$ to the data. At short integration times, $\Omega_z$ limited by additional noise contributions. For the vertical axis, the Allan deviation reaches a minimum of \SI{6}{\micro\radian/\second} at an integration time of approximately \SI{0.4}{\second}.}
    \label{fig:asdallan}
\end{figure*}
To address the self noise of the three-dimensional rotation sensor, Sagnac measurements are performed for each cavity without inducing a rotation to the optical table. Following Eq.~\ref{eqn:transformation}, this can be translated to a constant recorded rotation rate around each axis. The noise contributions can then be analyzed by computing the Welch amplitude spectral density \cite{Welch:1967}, as shown in Fig.~\ref{fig:asdallan}(a). At frequencies below \SI{1}{\hertz}, the dominant noise contribution in the $z$-axis (green trace in Fig.~\ref{fig:asdallan}(a)) is consistent with $1/f$-type flicker noise, as indicated by the approximately linear decrease in the amplitude spectral density in this regime. 
At higher frequencies, the system exhibits a frequency-independent white-noise behavior. Various noise contributions, including laser noise, acoustic and thermal perturbations of the cavities, as well as digitization and readout noise originating from the FPGA-based signal processing chain, are transferred to the Sagnac frequency readout through the PDH locking scheme used to control the AOM frequencies. A detailed investigation of the individual noise sources will be the subject of future work. A similar behavior is observed for the $y$-axis (orange trace in Fig.~\ref{fig:asdallan}(a)). However, in this case the flicker-noise-dominated region extends to approximately \SI{10}{\hertz}, as evidenced by the continued linear decrease of the spectral density up to this frequency. Beyond this point, white noise again becomes the dominant contribution. In contrast, the $x$-axis (blue trace in Fig.~\ref{fig:asdallan}(a)) exhibits a different low-frequency behavior. Below \SI{1}{\hertz}, the spectral slope deviates from a simple linear dependence, suggesting the presence of additional noise mechanisms. The steeper slope compared to the other axes is indicative of a $1/f^2$-type process, commonly referred to as Brownian noise. Such noise is often associated with stochastic fluctuations in environmental parameters, particularly temperature \cite{Ng1993}. At frequencies above \SI{1}{\hertz}, the spectrum of the $x$-axis is again dominated by white noise. The origin of the differing transition frequencies between flicker- and white-noise-dominated regimes across the three axes is not yet fully understood. One plausible explanation lies in the three-dimensional reconstruction algorithm. In the chosen coordinate system, the determination of rotations about the $z$-, $y$-, and $x$-axes requires one, two, and three cavities, respectively. This increasing complexity may lead to different noise propagation characteristics. In particular, the $x$-axis combines information from all cavities and is therefore more susceptible to accumulated noise contributions. This effect may be further amplified by the inclusion of the lowest-performing cavity, which has a finesse of only 25,000.

In addition, distinct noise peaks are observed on all axes. The peak at \SI{15}{\hertz} is of acoustic origin and is attributed to vibrations from the piezo. The \SI{50}{\hertz} peak arises from electronic interference. The feature at \SI{64}{\hertz} is again acoustically induced and can be traced back to the high-efficiency particulate air (HEPA) filter fan unit operating in the laboratory. Finally, the peak at \SI{77}{\hertz} is also of acoustic origin and is caused by the FPGA electronics used for PDH stabilization. This could be verified by performed accelerometer measurements. 

Furthermore, the classic Allan deviation \cite{Allan1966} is calculated to evaluate the performance of the rotation rate measurements, as shown in Fig.~\ref{fig:asdallan}(b). The performance of the gyroscopes is fundamentally limited by white noise at short integration times, as illustrated by the black dashed lines in Fig.~\ref{fig:asdallan}(b). These lines are obtained by fitting the function $\sigma_{\mathrm{wn}} = \frac{A}{\sqrt{\tau}}$ to the data, where $A$ represents the white-noise-limited sensitivity. In the vertical axis, larger deviations from purely white-noise-limited performance are observed, which are attributed to vibrations induced by the HEPA filter. The vertical signal is determined primarily by the horizontal cavity, which has the largest surface area exposed to airflow from the HEPA filter. The best sensitivity is reached at \SI{0.1}{\second} ($x$ and $y$-axis) with \SI{29}{\micro\radian/\second} and \SI{7}{\micro\radian/\second} respectively, whereas the vertical axis reaches its minimum at \SI{0.4}{\second} with \SI{6}{\micro\radian/\second}. In the regime of white-noise limited read-out, the sensitivity amounts to $A_x$=\SI{7}{\micro \radian / \second / \sqrthertz} ($x$-axis), $A_y$=\SI{2}{\micro \radian / \second / \sqrthertz} ($y$-axis), and $A_z$=\SI{3}{\micro \radian / \second / \sqrthertz} ($z$-axis). The achievable sensitivity differs between the axes, with the $x$-axis exhibiting the lowest performance. This behavior is already evident in Fig.~\ref{fig:asdallan}(a), where the $x$-axis is the only direction affected by additional Brownian noise contributions. A further factor contributing to this discrepancy is the reconstruction of the $x$-axis projection in the chosen coordinate system, which requires the rotation rate signals from all three cavities and therefore inherits their respective performance limitations, as stated earlier. In addition, the cavities are not optically symmetric and differ in finesse and incoupling efficiency for instance, which likely constitutes the dominant source of the observed performance variations between the axes.

One can follow from Ref.~\cite{U.Schreiber2013} that the theoretically achievable sensitivity of this setup would be in the \si{\nano\radian\per\second\per\sqrt{\hertz}} regime, assuming that the system is limited solely by white noise in the form of fundamental quantum noise (shot noise). We find that the stability of the prototype is currently limited by air movements and precise temperature stabilization. Temperature fluctuations and instabilities can affect the ambient pressure in the laboratory, leading to changes in the refractive index of the air traversed by the laser beams. This may represent a significant limitation. Addressing these aspects by enclosing the system in a vacuum environment, in combination with the implementation of improved super-polished cavity mirrors, could enhance the performance by orders of magnitude, potentially approaching the theoretical limit. Despite its compact size and the comparably small scale factor, the prototype achieves a sensitivity of a few \SI{}{\micro\radian/\second/\sqrthertz} without further optimization of the experimental design, demonstrating the potential of a three-dimensional free-space PRG system.

\section{Conclusion}

We have presented a transportable three-dimensional free-space PRG based on a tetrahedral cavity geometry and demonstrated its capability to reconstruct full vectorial ground rotations. The system achieves sensitivities in the \SI{}{\micro\radian/\second/\sqrthertz} regime in all three spatial dimensions and successfully resolves controlled rotational excitations mimicking seismic signals. The present performance may be enhanced through the implementation of a vacuum environment, active thermal stabilization, and improved mirror quality. Such modifications represent promising approaches to reaching the theoretically achievable sensitivity of the system. The use of the fourth cavity offers the prospect of reducing statistical uncertainties, lowering the effective noise floor, and identifying systematic effects. These results demonstrate the feasibility of compact three-dimensional free-space PRG systems for rotational seismology and geodesy. Compact and transportable instruments of this type constitute a promising step toward distributed rotational sensor arrays and may enable high-resolution local measurements of seismic wavefields at geographically diverse sites.

\begin{acknowledgments}
We acknowledge experimental support from J.~Spanier, J.~Gutsche, and P.~Stürmer, as well as fruitful discussions with K.~U.~Schreiber, H.~Igel, O.~Gerberding, K.~Isleif and all members of the RING collaboration.
\end{acknowledgments}

\section*{DATA AVAILABILITY STATEMENT}
The data cannot be made publicly available upon publication because no suitable repository exists
for hosting data in this field of study. The data that support the findings of this study are available upon
reasonable request from the authors
\section*{FUNDING}
We acknowledge financial support from the European Research Council ERC through grants No. 101123334 and 101213032.

\bibliography{literatur}

@article{G.Sagnac1914,
	author = {{Sagnac, G.}},
	title = {Effet tourbillonnaire optique. La circulation de l'éther lumineux dans un interférographe tournant},
	journal = {J. Phys. Theor. Appl.},
    url= "https://doi.org/10.1051/jphystap:019140040017700",
	year = 1914,
}

@article{U.Schreiber2013,
    author = {Schreiber, Karl Ulrich and Wells, Jon-Paul R.},
    title = {Invited Review Article: Large ring lasers for rotation sensing},
    journal = {Review of Scientific Instruments},
    volume = {84},
    number = {4},
    pages = {041101},
    year = {2013},
    month = {04},
    url = {https://doi.org/10.1063/1.4798216},
}

@article{A.Gebauer2020,
  title = {Reconstruction of the Instantaneous Earth Rotation Vector with Sub-Arcsecond Resolution Using a Large Scale Ring Laser Array},
  author = {Gebauer, Andr\'e and Tercjak, Monika and Schreiber, Karl Ulrich and Igel, Heiner and Kodet, Jan and Hugentobler, Urs and Wassermann, Joachim and Bernauer, Felix and Lin, Chin-Jen and Donner, Stefanie and Egdorf, Sven and Simonelli, Andrea and Wells, Jon-Paul R.},
  journal = {Phys. Rev. Lett.},
  volume = {125},
  issue = {3},
  pages = {033605},
  numpages = {5},
  year = {2020},
  month = {Jul},
  publisher = {American Physical Society},
  url = {https://link.aps.org/doi/10.1103/PhysRevLett.125.033605}
}

@article{Schreiber2023,
author={Schreiber, K. Ulrich
and Kodet, Jan
and Hugentobler, Urs
and Kl{\"u}gel, Thomas
and Wells, Jon-Paul R.},
title={Variations in the Earth's rotation rate measured with a ring laser interferometer},
journal={Nature Photonics},
year={2023},
url={https://doi.org/10.1038/s41566-023-01286-x}
}

@article{Schreiber2025,
author = {K. Ulrich Schreiber  and Urs Hugentobler  and Jan Kodet  and Simon Stellmer  and Thomas Klügel  and Jon-Paul R. Wells },
title = {Gyroscope measurements of the precession and nutation of Earth’s axis},
journal = {Science Advances},
year = {2025},
URL = {https://www.science.org/doi/abs/10.1126/sciadv.adx6634},
}

@article{Chen2025,
author = {Yuxuan Chen and Yuhong Zhong and Kui Liu and Yawen Liu and Yangsheng Cai and Zhanhao Liu and Lei Zheng and Zhiyuan Wang and Karl Ulrich Schreiber and Zehuang Lu and Jie Zhang and Liangcheng Tu and Jun Luo},
journal = {Opt. Lett.},
publisher = {Optica Publishing Group},
title = {Giant 64 m2 passive green laser gyroscope: solution in Earth rotation sensing and geophysical detection},
year = {2025},
url = {https://opg.optica.org/ol/abstract.cfm?URI=ol-50-12-3883},
}

@article{Feng2023,
author = {Xiaohua Feng and Kui Liu and Yuxuan Chen and Haobo Zhang and Zongyang Li and Fenglei Zhang and Karl Ulrich Schreiber and Zehuang Lu and Jie Zhang},
journal = {Appl. Opt.},
publisher = {Optica Publishing Group},
title = {Three-wave differential locking scheme in a 12-m-perimeter large-scale passive laser gyroscope},
year = {2023},
url = {https://opg.optica.org/ao/abstract.cfm?URI=ao-62-4-1109},
}

@article{Igel2021,
    author = {Igel, Heiner and Schreiber, Karl Ulrich and Gebauer, André and Bernauer, Felix and Egdorf, Sven and Simonelli, Andrea and Lin, Chin-Jen and Wassermann, Joachim and Donner, Stefanie and Hadziioannou, Céline and Yuan, Shihao and Brotzer, Andreas and Kodet, Jan and Tanimoto, Toshiro and Hugentobler, Urs and Wells, Jon-Paul R},
    title = {ROMY: a multicomponent ring laser for geodesy and geophysics},
    journal = {Geophysical Journal International},
    volume = {225},
    number = {1},
    pages = {684-698},
    year = {2021},
    month = {01},
    url = {https://doi.org/10.1093/gji/ggaa614}
}

@Book{Schreiber2023Book,
  author    = {Ultrich Schreiber and Jon-Paul Wells},
  title     = {Rotation Sensing with Large Ring Lasers: Applications in Geophysics and Geodesy},
  publisher = {Cambrige University Press},
  year      = {2023}
}

@article{brotzer2025,
    author = {Brotzer, Andreas and Igel, Heiner and Bernauer, Felix and Wassermann, Joachim and Kodet, Jan and Schreiber, Karl Ulrich and Zenner, Jannik and Stellmer, Simon},
    title = {On environment-related instrumental effects of ROMY (ROtational Motions in seismologY): A prototype, multi-component, heterolithic ring laser array},
    journal = {Review of Scientific Instruments},
    year = {2025},
    url = {https://doi.org/10.1063/5.0242127},
}

@Book{brotzerphd,
  author       = {Andreas Brotzer},
  title        = {The ROMY Ring Laser Array
for Seismology:
Instrumental Characterization,
Low Noise Limits for Rotations and
Seismological Applications},
  institution  = {University Munich},
publisher = {Dissertation},
  year         = {2024}
  }

@article{Igel2005,
author = {Igel, Heiner and Schreiber, Ulrich and Flaws, Asher and Schuberth, Bernhard and Velikoseltsev, Alex and Cochard, Alain},
title = {Rotational motions induced by the M8.1 Tokachi-oki earthquake, September 25, 2003},
journal = {Geophysical Research Letters},
volume = {32},
number = {8},
pages = {},
url = {https://agupubs.onlinelibrary.wiley.com/doi/abs/10.1029/2004GL022336},
year = {2005}
}

@article{Igel2011,
author = {Igel, Heiner and Nader, Maria-Fernanda and Kurrle, Dieter and Ferreira, Ana M. G. and Wassermann, Joachim and Schreiber, K. Ulrich},
title = {Observations of Earth's toroidal free oscillations with a rotation sensor: The 2011 magnitude 9.0 Tohoku-Oki earthquake},
journal = {Geophysical Research Letters},
volume = {38},
number = {21},
pages = {},
url = {https://agupubs.onlinelibrary.wiley.com/doi/abs/10.1029/2011GL049045},
year = {2011}
}

@article{Schreiber2003,
author = {Schreiber, K. Ulrich and Klügel, Thomas and Stedman, Geoffrey E.},
title = {Earth tide and tilt detection by a ring laser gyroscope},
journal = {Journal of Geophysical Research: Solid Earth},
volume = {108},
number = {B2},
pages = {},
url = {https://agupubs.onlinelibrary.wiley.com/doi/abs/10.1029/2001JB000569},
year = {2003}
}

@article{Liu:19,
author = {K. Liu and F. L. Zhang and Z. Y. Li and X. H. Feng and K. Li and Z. H. Lu and K. U. Schreiber and J. Luo and J. Zhang},
journal = {Opt. Lett.},
number = {11},
pages = {2732--2735},
publisher = {Optica Publishing Group},
title = {Large-scale passive laser gyroscope for earth rotation sensing},
volume = {44},
month = {Jun},
year = {2019},
url = {https://opg.optica.org/ol/abstract.cfm?URI=ol-44-11-2732},
}

@article{Korth_2016,
url = {https://doi.org/10.1088/0264-9381/33/3/035004},
year = {2016},
month = {jan},
publisher = {IOP Publishing},
volume = {33},
number = {3},
pages = {035004},
author = {Korth, W Z and Heptonstall, A and Hall, E D and Arai, K and Gustafson, E K and Adhikari, R X},
title = {Passive, free-space heterodyne laser gyroscope},
journal = {Classical and Quantum Gravity}
}

@article{Ezekiel1977,
    author = {Ezekiel, S. and Balsamo, S. R.},
    title = {Passive ring resonator laser gyroscope},
    journal = {Applied Physics Letters},
    volume = {30},
    number = {9},
    pages = {478-480},
    year = {1977},
    month = {05},
    issn = {0003-6951},
    url = {https://doi.org/10.1063/1.89455},
}

@article{Huang:24,
author = {Huimin Huang and Yujia Cao and Lanxin Zhu and Yanjun Chen and Wenbo Wang and Fangshuo Shi and Zhengbin Li},
journal = {CLEO 2024},
publisher = {Optica Publishing Group},
title = {A Giant interferometric fiber optic gyroscope with Low Self-Noise for Geophysical Rotation Sensing},
year = {2024},
url = {https://opg.optica.org/abstract.cfm?URI=CLEO_FS-2024-JTu2A.203},
}

@article{Zenner:26,
author = {Jannik Zenner and Karl Ulrich Schreiber and Simon Stellmer},
journal = {Optics Letters},
publisher = {Optica Publishing Group},
title = {Hänsch-Couillaud locking of a large Sagnac interferometer: advancing below the flicker floor},
year = {2026},
url = {https://opg.optica.org/ol/abstract.cfm?URI=ol-51-2-500},
}

@article{Zenner25,
    author = {Jannik Zenner and Karl Ulrich Schreiber and Simon Stellmer},
    title = {Stabilizing the free spectral range of a large ring laser },
    journal = {Optics Letters},
    year = {2025},
    url = {https://doi.org/10.1364/OL.550265}
    }

@article{Black2001,
    author = {Black, Eric D.},
    title = {An introduction to Pound–Drever–Hall laser frequency stabilization},
    journal = {American Journal of Physics},
    year = {2001},
    month = {01},
    url = {https://doi.org/10.1119/1.1286663}
    }

@manual{ADIS,
  author       = {{Analog Devices}},
  title        = {{ADIS16465} Precision MEMS IMU Module Data Sheet},
  organization = {Analog Devices},
  year         = {2023},
  url          = {https://www.analog.com/media/en/technical-documentation/data-sheets/adis16465.pdf},
  urldate      = {2026-01-07},
}

@manual{E15Laser,
  author       = {NKT Photonics},
  year         = {2024},
  title        = {Koheras ADJUSTIK Product Guide},
  organization = {NKT Photonics},
  url          = {https://www.nktphotonics.com/product-manuals-and-documentation/},
  urldate      = {2026-02-26}
}

@article{Welch:1967,
  author={Welch, P.},
  journal={IEEE Transactions on Audio and Electroacoustics}, 
  title={The use of fast Fourier transform for the estimation of power spectra: A method based on time averaging over short, modified periodograms}, 
  year={1967},
  volume={15},
  number={2},
  pages={70-73},
  doi={10.1109/TAU.1967.1161901}
}

@Book{Ng1993,
  author       = {Ng, L C},
  title        = {On the application of Allan variance method for Ring Laser Gyro performance characterization},
  publisher  = {Lawrence Livermore National Lab., CA (United States)},
  url          = {https://www.osti.gov/biblio/10196087},
  year         = {1993}
  }

@article{Allan1966,
  author={Allan, D.W.},
  journal={Proceedings of the IEEE}, 
  title={Statistics of atomic frequency standards}, 
  year={1966},
  volume={54},
  number={2},
  pages={221-230},
  doi={10.1109/PROC.1966.4634}}

@article{Ghufran2024,
title = {True north measurement: A comprehensive review of Carouseling and Maytagging methods of gyrocompassing},
journal = {Measurement},
volume = {226},
pages = {114121},
year = {2024},
url = {https://www.sciencedirect.com/science/article/pii/S0263224124000058},
author = {Ghufran Aqeel Asif and Nur Hazliza Ariffin and Norazreen Ab Aziz and Mohd Hadri Hafiz Mukhtar and Norhana Arsad}
}

@article{Brotzer2023,
    author = {Brotzer, Andreas and Igel, Heiner and Stutzmann, Eléonore and Montagner, Jean‐Paul and Bernauer, Felix and Wassermann, Joachim and Widmer‐Schnidrig, Rudolf and Lin, Chin‐Jen and Kiselev, Sergey and Vernon, Frank and Schreiber, Karl Ulrich},
    title = {Characterizing the Background Noise Level of Rotational Ground Motions on Earth},
    journal = {Seismological Research Letters},
    volume = {95},
    number = {3},
    pages = {1858-1869},
    year = {2023},
    month = {12},
    url = {https://doi.org/10.1785/0220230202}
}

@article{Virgilio2024,
  title = {Noise Level of a Ring Laser Gyroscope in the Femto-Rad/s Range},
  author = {Di Virgilio, Angela D. V. and Bajardi, Francesco and Basti, Andrea and Beverini, Nicol\`o and Carelli, Giorgio and Ciampini, Donatella and Di Somma, Giuseppe and Fuso, Francesco and Maccioni, Enrico and Marsili, Paolo and Ortolan, Antonello and Porzio, Alberto and Vitali, David},
  journal = {Phys. Rev. Lett.},
  year = {2024},
  publisher = {American Physical Society},
  url = {https://link.aps.org/doi/10.1103/PhysRevLett.133.013601}
}

@article{Virgilio2019,
author = {Virgilio, A. and Beverini, N. and Carelli, G. and Ciampini, Donatella and Fuso, Francesco and Maccioni, Enrico},
year = {2019},
title = {Analysis of ring laser gyroscopes including laser dynamics},
journal = {The European Physical Journal C},
doi = {10.1140/epjc/s10052-019-7089-5}
}

@article{Capezzuto2024,
author = {Capezzuto, Marialuisa and Gaudiosi, Guido and Nardone, Lucia and D’Alema, Ezio and D’Ambrosio, Davide and Manzo, Roberto and Giorgini, Antonio and Malara, Pietro and Natale, Paolo and Gagliardi, Gianluca and Amato, Luigi and Galluzzo, Danilo and Avino, Saverio},
year = {2024},
title = {Fiber-optic gyroscope for rotational seismic ground motion monitoring of the Campi Flegrei volcanic area},
journal = {Applied Optics},
doi = {10.1364/AO.518354}
}

@article{Simonelli2018,
    author = {Simonelli, A and Igel, H and Wassermann, J and Belfi, J and Di Virgilio, A and Beverini, N and De Luca, G and Saccorotti, G},
    title = {Rotational motions from the 2016, Central Italy seismic sequence, as observed by an underground ring laser gyroscope},
    journal = {Geophysical Journal International},
    year = {2018},
    url = {https://doi.org/10.1093/gji/ggy186}
}

@article{Chung2022,
  title    = "Seismic waves of the 10 December 2020 M6.6 Yilan earthquake
              observed by interferometric fiber-optic gyroscope",
  author   = "Chung, Hung-Pin and Chang, Sheng-Han and Hsieh, Ching-Lu and
              Yang, Hsuan and Chen, Chii-Chang and Liu, Jann-Yenq and Yen,
              Horng-Yuan and Chen, Yen-Hung",
  journal  = "Terrestrial, Atmospheric and Oceanic Sciences",
  year     =  2022
}

@article{Yuan2020,
    author = {Yuan, Shihao and Simonelli, Andreino and Lin, Chin‐Jen and Bernauer, Felix and Donner, Stefanie and Braun, Thomas and Wassermann, Joachim and Igel, Heiner},
    title = {Six Degree‐of‐Freedom Broadband Ground‐Motion Observations with Portable Sensors: Validation, Local Earthquakes, and Signal Processing},
    journal = {Bulletin of the Seismological Society of America},
    year = {2020},
    url = {https://doi.org/10.1785/0120190277},
}

@article{BrotzerIgel2025, title={On single-station, six degree-of-freedom observations of local to regional seismicity at the Piñon Flat Observatory in Southern California}, volume={4}, url={https://seismica.library.mcgill.ca/article/view/1416},journal={Seismica}, author={Brotzer, Andreas and Igel, Heiner and Bernauer, Felix and Wassermann, Joachim and Mellors, Robert and Vernon, Frank}, year={2025}}

@Article{Izgi2021,
AUTHOR = {Izgi, Gizem and Eibl, Eva P. S. and Donner, Stefanie and Bernauer, Felix},
TITLE = {Performance Test of the Rotational Sensor blueSeis-3A in a Huddle Test in Fürstenfeldbruck},
JOURNAL = {Sensors},
YEAR = {2021},
URL = {https://www.mdpi.com/1424-8220/21/9/3170},
DOI = {10.3390/s21093170}
}

\end{document}